\documentstyle[11pt,paspconf,epsf]{article}

\def\edcomment#1{\iffalse\marginpar{\raggedright\sl#1\/}\else\relax\fi}
\reversemarginpar

\begin {document}

\title{Extrasolar Planets: A Galactic Perspective}

\author{I. Neill Reid}%

\affil{Space Telescope Science Institute, 3700 San Martin Drive, 
Baltimore, MD 21218, USA}




\begin{abstract}

The host stars of extrasolar planets tend to be metal-rich. We have examined the data for these stars for evidence of trends in other galactic parameters, without success. However, several ESP hosts are likely to be members of the thick disk population, indicating that planet formation has occurred throughout the full lifetime of the Galactic disk. We briefly consider the radial metallicity gradient and age-metallicity relation of the Galactic disk, and complete a back-of-the envelope estimate of the likely number of solar-type stars with planetary companions with $6 < R < 10$ kpc. 

\end{abstract}

\section {Introduction}

Exploitation of major scientific discoveries tends to follow a familiar 
pattern. Immediately following the initial discovery, the main focus is 
on verification, testing the initial analysis against alternative 
explanations for the same phenomena. The second phase centres on 
consolidation, acquiring additional observations and/or improved 
theoretical data on particular phenomena. The third phase is reached 
with the discovery of sufficient observational examples to map out a 
substantial fraction of the phase space occupied by key parameters.

After only a decade, investigations of extrasolar planets have clearly 
entered the third stage. More than 145 planetary-mass companions are 
known around over 125 main sequence dwarfs and red giants. The overall 
properties of these planets are discussed elsewhere in this volume (see 
the reviews by Marcy and Mayor). It is now well established that the 
frequency of (currently-detectable) planetary systems increases with 
increasing metallicity of the parent star (see Valenti, this 
conference). Here, we concentrate on analysing other salient properties 
of the extrasolar planetary (ESP) host stars, looking for evidence of 
potential correlations. 

\section {Wide planetary-mass companions of low-mass dwarfs}

Recent months have seen the imaging of low-mass companions to several 
young stars in the Solar Neighbourhood. Neuhauser et al (2005) have 
identified a faint, common proper motion (cpm) companion of GQ Lupi, a 
$\sim0.45 M_\odot$ member of the Lupus I group ($\tau \sim 1-2$Myrs); 
the companion lies at a separation of $\sim100$ AU and has a likely 
mass between 1 and 42 Jupiter masses (M$_J$). Similarly, Chauvin et al 
(2005b) have discovered a faint cpm companion to AB Pic, a K2 dwarf, 
member of the $\sim30$ Myr-old Tucana-Horologium association; the 
companion is $\sim245$ AU distant, with a likely mass of 13-14 M$_J$. 
Finally, and most intriguingly, Chauvin et al (2004a) have shown that 
2MASS J1207334-393254, a $\sim35 M_J$ brown dwarf member of the 
$\sim10$-Myr-old TW Hydrae association, has a wide cpm companion, 
separation $\sim60$ AU, with a likely mass of only 2-5 $M_J$ (see also 
Schneider, this conference).

\begin{figure}
\plotfiddle{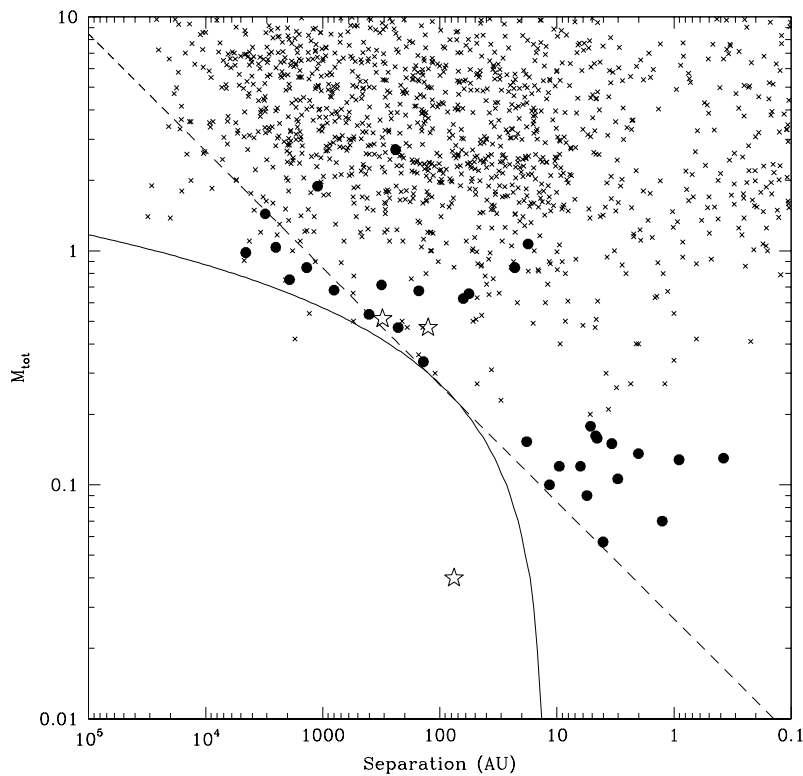}{8 truecm}{0}{55}{55}{-190}{-90}
\caption{Wide planetary-mass companions of low-mass dwarfs: total mass as a function of separation in binary systems (adapted from Burgasser et al, 2003). The crosses plot data for stellar binaries, while solid points identify systems with brown dwarf components; the solid line outlines a linear relation between maximum separation and total mass, while the dotted line plots $a_{\max} \propto M_{tot}^2$. The stars mark the location of 2M1207AB, GQ Lupi and AB Pic AB.}
\end{figure}

Should we classify these low-mass companions as planets or brown 
dwarfs? In my opinion, all of these objects are brown dwarfs. The mass 
estimates themselves, based on evolutionary models, offer no 
discrimination, since, while they overlap with the planetary r\'egime, 
it remains unclear whether they lie beyond a lower limit to the brown 
dwarf mass spectrum (if such exists). The large distance between each 
companion and its primary is a more telling parameter. There is little 
question that the Solar System planets formed from the Sun's 
protoplanetary disk; it seems reasonable to apply the same criterion in 
categorising extrasolar planets. This test offers no discriminatory 
power at small separations, such as with the 11$M_J$ companion of HD 
114762, which has an orbital semi-major axis of 0.3 AU. However, one 
might argue that disks are likely to have problems forming massive 
companions at large radii. Thus, it seems unlikely that a 35$M_J$ brown 
dwarf would possess a protoplanetary disk sufficiently massive that it 
could form a 2-5 $M_J$ companion at a distance of 60 AU from the 
primary.

Nonetheless, this debate over terminology should not be allowed to 
obscure the significance of these discoveries, particularly 2M1207AB. 
Recent analyses show that the maximum separation of binaries in the 
field appears to correlate with the total mass (Burgasser et al, 2003). 
The newly discovered systems lie at the extremes of this distribution 
(Figure 1). 2M1207AB is particularly notable, with not only a projected 
separation, $\Delta$, four times higher than comparable brown dwarf 
binaries, but also a very low mass ratio, $q = {M_2 \over M_1} < 0.14$. 
All of these systems would be extremely difficult to detect at ages 
exceeding$\sim10^8$ years; thus, their absence among field binaries 
could reflect either dynamical evolution (and system disruption), or 
observational selection effects. 

\section { The planetary host stars}

The overwhelming majority of planetary systems have been discovered 
through radial velocity surveys, with a handful of recent additions 
from transit surveys and microlensing programs. The radial velocity 
surveys focus on solar-like stars, for obvious reasons, so it is not 
surprising that most of the ESP hosts are late-F to early-K dwarfs; 
moreover, almost all lie within 50 parsecs of the Sun (Figure 2). 

\begin{figure}
\plotfiddle{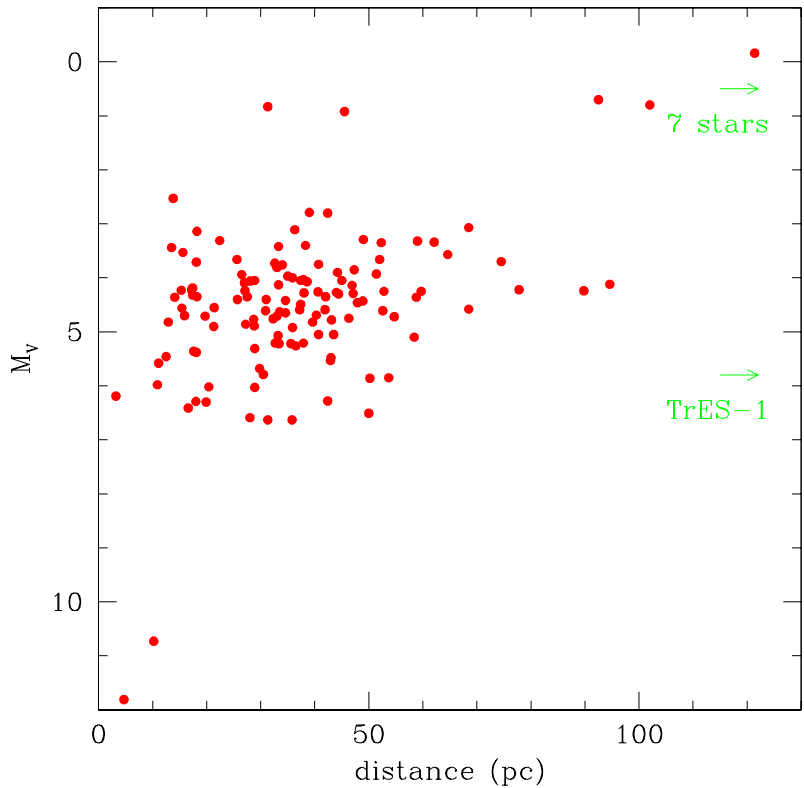}{8 truecm}{0}{55}{55}{-190}{-90}
\caption{The distance distribution of stars known to have planetary mass companions; absolute magnitudes are primarily based on Hipparcos parallax data.}
\end{figure}

Approximately 90\% of the solar-type stars within 25 parsecs of the Sun 
are included in either the UC or Geneva surveys, although only the UC 
survey has published a catalogue of non-detections (Nidever et al, 
2001). Building on the local completeness of these surveys, Figure 3 
shows the distribution of semi-major axes/projected separation as a 
function of mass for all known companions of solar-type stars within 25 
parsecs of the Sun; the distribution 
clearly shows the brown dwarf desert, and provides the most effective 
demonstration that extrasolar planets are not simply a low-mass tail to 
the stellar/brown dwarf companion mass function.

\begin{figure}
\plotfiddle{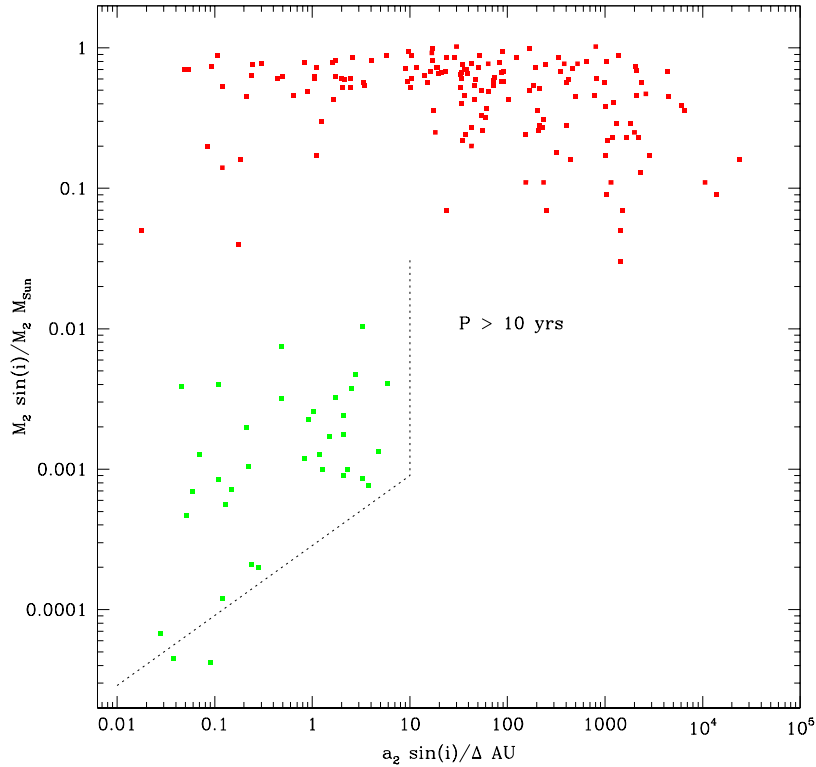}{8 truecm}{0}{55}{55}{-190}{-90}
\caption{Companions mass as a function of separation for solar-type stars within 25 parsecs of the Sun; the dotted lines mark the effective limits of the planetary radial velocity surveys. The scarcity of systems with masses betwee 0.01 and 0.1 $M_\odot$ is the brown dwarf desert.}
\end{figure}

\subsection {Stars and planetary properties}

Given the correlation between planetary frequency and metallicity, one 
might expect a bias towards higher mass planets in metal-rich stars; 
the current data, however, do not support that contention. This 
suggests that high metallicity acts as a trigger for planet formation, 
rather than playing a key role in the formation mechanism itself. 
Metal-rich systems ([m/H]$>0.1$) do include a higher proportion of 
short-period, hot Jupiters, perhaps reflecting higher viscous drag in 
the protoplanetyary disk (Sozzetti, 2004; Boss, this conference). 

There is no obvious direct correlation between the mass of the primary 
star and the masses of planetary companions. Thus, the 
$\sim0.35M_\odot$ M3 dwarf, Gl 876, has two planets with masses 
comparable to Jupiter. On the other hand, one might expect an upper 
limit to the mass distribution to emerge, simply because lower mass 
stars are likely to have lower mass protoplanetary disks (see also 
Marcy, this conference).

\subsection {Metallicities and the thick disk}

Chemical abundance, particularly individual elemental abundance ratios, 
serves as a population discriminant for the ESP host stars. Halo stars 
have long been known to possess $\alpha$-element abundances (Mg, Ti, O, 
Ca, Si) that are enhanced by a factor of 2-3 compared to the Sun. This 
is generally attributed to the short formation timescale of the halo 
(Matteucci \& Greggio, 1983): $\alpha$-elements are produced by rapid 
$\alpha$-capture, and originate in Type II supernovae, massive stars 
with evolutionary lifetimes of 10$^7$ to 10$^8$ years. In contrast, 
Type I supernovae, which are produced by thermal runaway on an 
accreting white dwarf in a binary system, have evolutionary timescales 
of 1-2 Gyrs. These systems produce a much higher proportion of Fe; 
thus, their ejecta drive down the [$\alpha$/Fe] ratio in the ISM, and 
in newly forming stars.

Recent high-resolution spectroscopic analyses of nearby high velocity 
stars provide evidence that the thick disk is also $\alpha$-enhanced 
(Fuhrmann, 1998, 2004; Prochaska, 2000). The thick disk is the extended 
population originally identified from polar star counts by Gilmore \& 
Reid (1983); current theories favour an origin through dynamical 
excitation by a major merger early in the history of the Milky Way 
(ref). With abundances in the range $-1 < [m/H] < -0.3$ (Figure 4, upper panel), 
the thick disk clearly formed after the Population II halo, but before 
Type I supernovae were able to drive up the iron abundance. Thus, the 
thick disk population almost certainly comprises stars from the 
original Galactic disk, which formed within the first 1-2 Gyrs of the 
Milky Way's history.

\begin{figure}
\plotfiddle{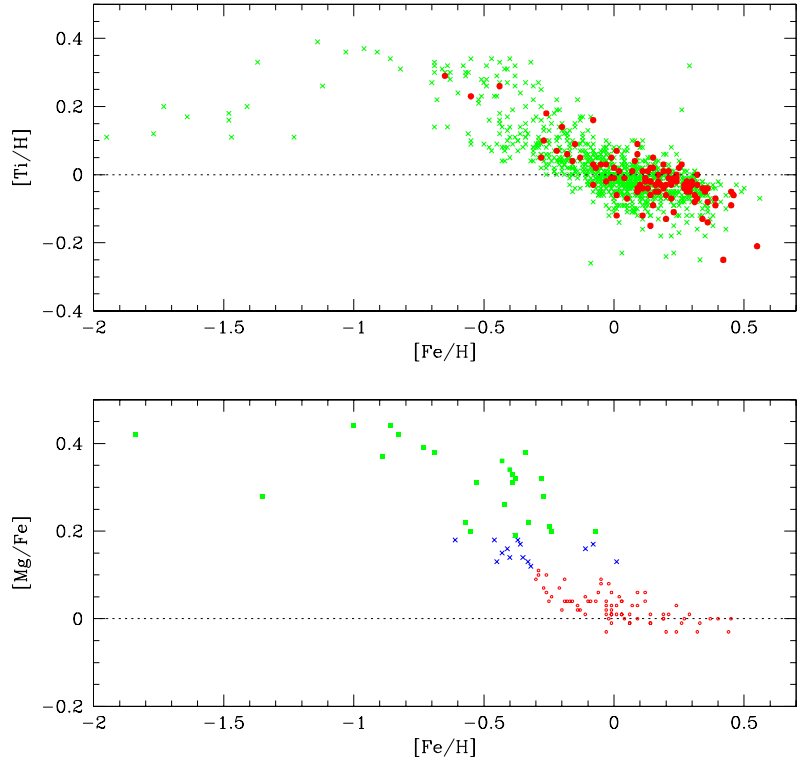}{8 truecm}{0}{55}{55}{-190}{-90}
\caption{$\alpha$-element abundances in nearby stars. The lower panel plots data for nearby stars from Fuhrmann (1998), where the solid triangles mark stars identified as members of the thick disk. The upper panel plots data from Valenti \& Fischer's (2005) analysis of stars in the Berkeley/Carnegie planet survey; the solid points mark stars known to have planetary companions. Three stars (identified in the text) are almost certainly members of the thick disk.}
\end{figure}

What is the relevance of these observations to planet formation? 
Valenti \& Fischer (2005) have recently completed abundance analysis of 
the high-resolution spectroscopic data acquired by the Berkeley/Carnegie 
radial-velocity survey. Their analysis includes measurement of the abundance of 
Ti, an $\alpha$-element. The lower panel of Figure 4 shows the distribution of the full sample as a 
function of [Fe/H], identifying stars known to have planets. Three of 
the latter stars have $\alpha$-abundances consistent with thick disk 
stars (HD 6434, 0.48$M_J$ planet; HD 37124, 0.75$M_J$; HD 114762, 
11$M_J$), while three other stars (HD 114729, 0.82$M_J$; $\rho$ CrB, 
1.04$M_J$; HD 168746, 0.23$M_J$) have intermediate values of 
[$\alpha$/Fe]. These results strongly suggest that, even though planets 
may be extremely rare among metal-poor halo stars (Gilliland et al, 
2000), planetary systems have been forming in the Galactic disk since 
its initial formation.

\subsection {Kinematics}

Stellar kinematics are usually characterised using the Schwarzschild 
velocity ellipsoid; probability plots (Lutz \& Upgren, 1981) allow one 
to compare stellar samples that include multiple components with 
Gaussian velocity distributions (see, for example, Reid, Gizis \& 
Hawley, 2002). With the completion of the Geneva-Copehnhagen
survey of the Solar Neighbourhood  (Nordstr\"om et al., 2004), we 
have distances, proper motions, radial velocities and 
abundances\footnote{One should note that the abundances in this catalogue
are tied to 
the Sch\"uster et al {\sl uvby}-based metallicity scale, which has 
colour-dependent systematic errors (Haywood, 2002). Those systematic 
errors can be corrected.} for most solar-type stars within 40 
parsecs of the Sun. There are 1273 stars within that distance limit 
with $0 < (b-y) < 0.54$ and M$_V \ge 4.0$; analysing their kinematics, 
we have 
\begin{equation}
(U, V, W; \sigma_U, \sigma_V, \sigma_W) \qquad = \qquad (-9.6, -20.2, -
7.6; 38.8, 31.0, 17.0 \quad {\rm km s}^{-1})
\end{equation}
with an overall velocity dispersion, $\sigma_{tot} = 52.7$ kms$^{-1}$. 

There are 129 ESP hosts with accurate distances and space motions, and 
their mean kinematics are
\begin{equation}
(U, V, W; \sigma_U, \sigma_V, \sigma_W) \qquad = \qquad (-4.0, -25.5, -
20.4; 37.7, 22.9, 20.4 \quad {\rm km s}^{-1})
\end{equation}
with an overall velocity dispersion, $\sigma_{tot} = 48.6$ kms$^{-1}$.
All three stars identified as likely members of the thick disk (and two of the possible members) have high velocities (60 to 110 kms$^{-1}$) relative to the Sun. 

\begin{figure}
\plotfiddle{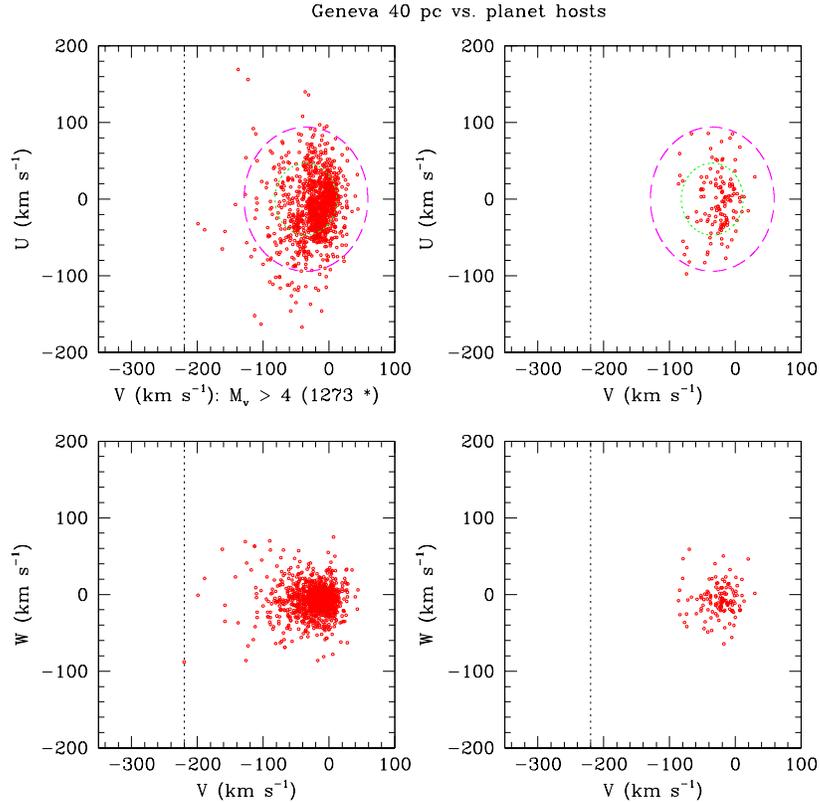}{8 truecm}{0}{55}{55}{-190}{-90}
\caption{The velocity distribution of ESP host stars (right hand panels) compared with the velocity distribution of solar-type stars within 40-parsecs of the Sun.}
\end{figure}

The two velocity distributions are very similar. The total velocity 
dispersion of the ESP hosts is slightly lower than the field, mainly 
reflecting the higher proportion of old, metal-poor stars in the latter 
population (this also accounts for the lower value of $\sigma_V$ for 
the ESP hosts). Interpreted in terms of the standard stellar diffusion 
model, $\sigma_{tot} \propto \tau^{1/3}$, the observed difference 
formally corresponds to a difference of only 20\% in the average age. 
The relatively high mean motion perpendicular to the Plane is somewhat
surprising, and may warrant further investigation. With that possible exception,
however, the velocity distribution of ESP host stars is not particularly unusual, 
given the underlying metallicity distribution.

\section {Planetary cartography in the Galaxy}

The ultimate goal of these statistical analyses is to answer a simple 
question: How many stars in the Galaxy are likely to have associated 
planetary systems? While we are still far from being able to construct 
a reliable detailed model for the stellar populations in the Milky Way, 
there have been some initial studies, notably by Gonzalez, Brownlee \& 
Ward (2001). Here, we briefly consider those models.

\subsection {Abundance gradients and the Galactic Habitable Zone}

Planetary habitability is likely to depend on many factors: distance 
from the parent star; the stellar lifetime; (perhaps) planetary mass, which may
depend on the metallicity of the system; (perhaps) the presence of a 
massive satellite; the existence of plate tectonics; and (perhaps) the 
space motion of the system. In contrast, from the perspective of planet 
formation, the correlation between planetary frequency and metallicity 
is the only significant factor that has emerged from statistical 
analysis of the known ESP systems. Extending the present results beyond 
the Solar Neighbourhood requires modeling of both the radial abundance 
gradient and the age-metallicity relation of the Galactic disk.

Most studies assume a logarithmic radial abundance gradient; for 
example, Gonzalez et al (2001) adopt a gradient, ${\delta [m/H]} \over {\delta R}$, of 
-0.07 dex kpc$^{-1}$. More recent analyses of intermediate-age Cepheids 
(Andrievsky et al, 2002) suggest a more complex radial distribution, with a
flatter gradient in the vicinity of the Solar Radius, but steeper gradients 
outwith these limits. Figure 5a matches both distributions against 
[O/H] abundances for HII regions (from Shaver et al, 1983); we also show
current estimates for the Sun, and for the Hyades, Pleiades and Praesepe clusters.
By and large, the data favour the complex gradient, but one should note that 
extrapolating the
inner gradient implies unreasonably high metallicities at $R < 4$kpc. It
is probably more reasonable to infer that we require more observations of
metallicity tracers in the inner Galaxy.

\begin{figure}
\plotfiddle{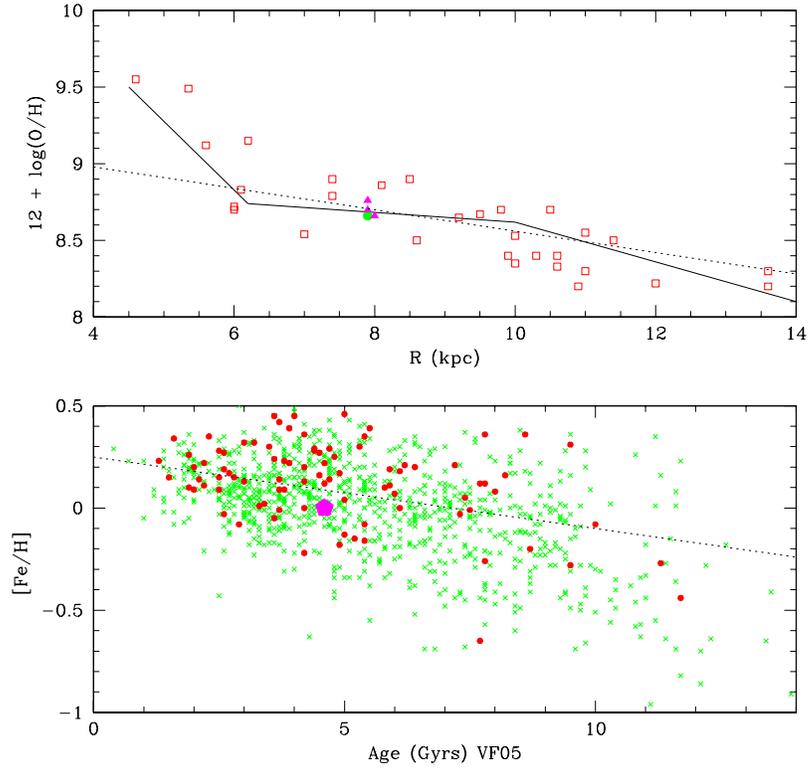}{8 truecm}{0}{55}{55}{-190}{-90}
\caption{The upper panel plots oxygen abundance data for HII regions as a function of Galactic radius (open squares, data from Shaver et al, 1983). We also show the oxygen abundances for the Sun (solid square) and for the three nearest open clusters, the Hyades, Pleiades and Praesepe (solid triangles). The dotted line plots the abundance gradient adopted by Gonzalez et al (2001); the solid line marks the composite gradient derived by Andrievsky et al (2002) from Cepheid data. The lower panel plots the age-metallicity distribution derived by Valenti \& Fischer (2005) for stars in the Berkeley/Carnegie radial velocity survey; solid points mark stars known to be ESP hosts. The pentagon marks the Sun.}
\end{figure}

The metallicity gradients plotted in the upper panel of Figure 6 are based on young objects - even
the Cepheids are less the 10$9$ years old. Thus, the distributions are
characteristic of the average metallicity of stars forming in the 
present-day Galaxy. Clearly, there is a substantial distribution of metallicity
among the stars in the Solar Neighbourhood, and one would expect comparable
dispersions at other radii. Gonzalez et al (2001) address this issue by
assuming an age-metallicity relation, ${\delta [m/H]} \over {\delta \tau}$
of 0.035 dex Gyr$^{-1}$; they assume a small dispersion in
metallicity ($<0.08$ dex). Note that this type of relation corresponds to a 
constant fractional increase in metallicity with time; that, in turn, 
requires either an increasing yield with increasing metallicity or an 
increasing star formation rate at a constant yield. 

The lower panel of Figure 6 matches the
Gonzalez et al age-metallicity relation against empirical results from
Valenti \& Fischer's (2005) analysis of stars in the Berkeley/Carnegie survey;
both ESP hosts and the Sun are separately identified. Unlike the classic
Edvardsson et al (1993) analysis, there is a trend in mean abundance with age. Overall, the observations indicate a much
broader dispersion in metallicity, at all ages, than assumed by Gonzalez et al
(2001)\footnote{Note that the Sun, which lies at the mode of the local abundance distribution, is somewhat metal rich for its age.}. As with the radial abundance gradient, further analysis is required
before settling on reliable values of these parameters.

\subsection {Planetary statistics in the Solar Circle}

Given these caveats, what can we say about the likely distribution of planetary systems in the Milky Way? For the moment, any calculations must be restricted to regions relatively close to the Sun. However, Figure 6 suggests that the local metallicity distribution may well be representative of stars with Galactic radii between 6 and 10 kiloparsecs. Taking this as a working assumption, we can calculate a back-of-the-envelope estimate of how many solar-type stars in this annulus might have planetary companions.

In making this calculation, we identify solar-type stars as stars with luminosities $4 < M_v < 6$; we assumed a density law
\begin{equation}
\rho(R) \quad = \quad \rho_0 e^{(R-R_0)/h} e^{-z/z_0}
\end{equation}
where $\rho_0 = 4.4 \times 10^{-3}$ stars pc$^{-3}$, with 90\% of the stars assigned to the disk and 10\% to the thick disk; $h$=2500 pc; and $z_0$=300 pc for the disk, and $z_0$=1000 pc for the thick disk. We assume an overall planetary frequency of 6\%.

\begin{figure}
\plotfiddle{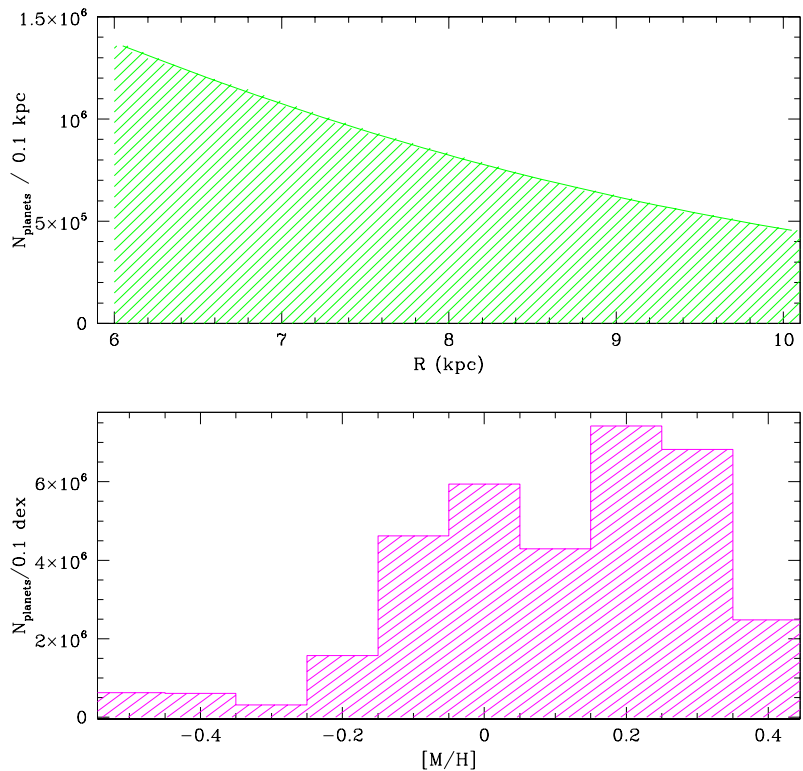}{8 truecm}{0}{55}{55}{-190}{-90}
\caption{The upper panel plots the expected numbers of ESP hosts as a function of Galactocentric distance for $6 < R < 10$ kpc; the lower panel plots the corresponding predicted metallicity distribution.}
\end{figure}

Given those starting assumptions, Figure 7 shows the expected radial density distribution and the metallicity distribution of ESP hosts. In total, we predict that $\sim3.5 \times 10^7$ solar-type stars are likely to have gas giant planetary companions with $a < 4$ AU. The space density increases with decreasing Galactocentric radius (density wins over surface area), and the metallicity distribution is approximately flat for $-0.1 < [m/H] < 0.3$. ESP hosts are not uncommon in the Solar Circle.

\section{Summary and conclusions}

More than 130 stars are now known to harbour planetary systems. The correlation between planetary frequency and metallicity is now well established, and probably reflects nature rather than nurture. We have analysed the statistical properties of the current sample of ESP hosts, looking for other possible trends and biases. With the possible exception of a higher mean velocity perpendicular to the Plane, the planetary hosts appear to be unremarkable members of the Galactic Disk.

Several ESP host stars show enhanced $\alpha$-element abundances and high velocities relative to the Sun, strongly suggesting that they are members of the thick disk. If so, this indicates that planet formation has been a constant presence in the Galactic Disk.

We briefly considered the radial abundance gradient and age-metallicity distribution of disk stars. Recent results suggest that the gradient is flatter in the vicinity of the Sun than the canonical value, but probably steepens at larger radii; there is relatively little reliable data for the inner disk, $R < 6$ kpc. The age-metallicity distribution of the Berkeley/Carnegie RV sample shows substantial dispersion at all ages. While the uncertainties in these parameters limit our ability to model the full Galaxy, we can use local statistics to estimate the planetary population in the vicinity of the Sun. We estimate that there are over $3.5 \times 10^7$ `RV-detectable' planetary systems with Galactocentric radii in the range 6 to 10 kpc. 

\begin{acknowledgements}
Thanks to Jeff Valenti for access to computer-readable tables listing the results of the spectroscopic analysis of the Berkeley/Carnegie radial velocity sample. 
\end{acknowledgements}

\end{document}